\documentclass[aps,pra,twocolumn,aps,floatfix,showpacs,tightenlines,showkeys,superscriptaddress,amsmath,amssymb]{revtex4-1}
\usepackage{amssymb,amsbsy,epsfig,color,graphicx}
\usepackage{color}
\usepackage{subfigure}
\usepackage{[longtable}
\usepackage{array}
\usepackage{dcolumn}   
\usepackage{cellspace}

\usepackage{etoolbox} 
\usepackage{lipsum} 
\usepackage[capitalize]{cleveref}

\makeatletter
\appto{\appendix}{%
	\@ifstar{\def\theequation@prefix{A.}}%
	{}%
}
\makeatother

\begin{document}


\title{Exact non-Markovian dynamics of Gaussian quantum channels: finite-time and asymptotic regimes}

\author{G. Torre}
\affiliation{Dipartimento di Ingegneria Industriale, Universit\`a degli Studi di Salerno,
Via Giovanni Paolo II, I-84084 Fisciano (SA), Italy}
\affiliation{INFN - Istituto Nazionale di Fisica Nucleare, Sezione di Napoli, Gruppo Collegato di Salerno,
	Via Giovanni Paolo II, I-84084 Fisciano (SA), Italy}

\author{F. Illuminati}
\thanks{Corresponding author: filluminati@unisa.it\\}
\affiliation{Dipartimento di Ingegneria Industriale, Universit\`a degli Studi di Salerno,
Via Giovanni Paolo II, I-84084 Fisciano (SA), Italy}
\affiliation{INFN - Istituto Nazionale di Fisica Nucleare, Sezione di Napoli, Gruppo Collegato di Salerno,
Via Giovanni Paolo II, I-84084 Fisciano (SA), Italy}

\date{April 9, 2018}

\begin{abstract}
We investigate the Markovian and non-Markovian dynamics of Gaussian quantum channels, exploiting a recently introduced necessary and sufficient criterion and the ensuing measure of non-Markovianity based on the violation of the divisibility property of the dynamical map. We compare the paradigmatic instances of Quantum Brownian motion (QBM) and Pure Damping (PD) channels, and for the former we find that the exact dynamical evolution is always non-Markovian in the finite-time as well as in the asymptotic regimes, for any nonvanishing value of the non-Markovianity parameter. If one resorts to the rotating wave approximated (RWA) form of the QBM, that neglects the anomalous diffusion contribution to the system dynamics, we show that such approximation fails to detect the non-Markovian nature of the dynamics. Finally, for the exact dynamics of the QBM in the asymptotic regime, we show that the quantifiers of non-Markovianity based on the distinguishability between quantum states fail to detect the non-Markovian nature of the dynamics.
\end{abstract}


\pacs{03.67.Hk, 42.50.Pq}

\maketitle


\section{Introduction}\label{Intro}

Memory effects in the dynamics of open quantum systems play a crucial role in various physical phenomena, from quantum biology~\cite{Lambert(2013), Thorwart2009234, Huelga2013} to quantum cryptography~\cite{PhysRevA.83.042321}, quantum metrology~\cite{PhysRevLett.109.233601}, and quantum control~\cite{PhysRevA.85.032321}. In the context of continuous variable (CV) systems, it has been shown that non-Markovianity of the dynamics can be usefully exploited for the enhancement of quantum teleportation protocols~\cite{PhysRevA.84.034305} and quantum cryptography tasks~\cite{PhysRevA.83.042321}. Consequently, great effort has been devoted to qualify and quantify the non-Markovianity content in the dynamics of open quantum systems (see~\cite{0034-4885-77-9-094001, RevModPhys.88.021002} for recent reviews). For the infinite-dimensional case and CV systems, important progress has been achieved in recent years, with the introduction of necessary and sufficient criteria for the non-Markovianity of Gaussian channels and non-Markovianity witnesses~\cite{PhysRevLett.115.070401, PhysRevLett.118.050401,  Groblacher(2015), PhysRevA.92.052122}.

Gaussian channels play a key role in the description of the open dynamics of quantum optical system~\cite{GaussianChannel}.
As such, it is important to characterize their non-Markovianity properties, possibly investigating their exact dynamics both in the finite and asymptotic time regimes.
Recently, we have introduced a necessary and sufficient criterion and a measure of non-Markovianity of bosonic Gaussian channels that are based on the violation of the divisibility property of the dynamical map~\cite{PhysRevLett.115.070401}. The measure has been applied to characterize the non-Markovianity of the time evolution within the rotating wave (RWA) approximation, that is typically employed in the description of the system dynamics in the weak coupling limit.

On the other hand, it is known that these approximations, for instance
in the case of the spin-boson model, fail to detect the non-Markovianity of the dynamics~\cite{PhysRevA.88.052111}.
Since a properly defined quantifier of non-Markovianity would allow to assess whether and to what extent these approximations describe the time evolution correctly, it is important to investigate the exact dynamics of Gaussian channels in the general case, making no approximations except the weak coupling limit assumption that preserves the form of the master equation for the dynamical map.
Aiming then at an exact characterization of the non-Markovianity of the dynamics, the violation of the divisibility condition of the dynamical map plays a crucial role. Indeed, as shown in Ref.~\cite{Groblacher(2015)}, both theoretically and experimentally, there is a close connection between the violation of the divisibility of the dynamical map and the structure of the system-environment interaction.

In the present paper we investigate the non-Markovianity of bosonic Gaussian channels in general terms for the exact channel dynamics without approximations, and we show that it persists even when the output state of the system evolving in the channel is independent of the input, i.e., when the system undergoes thermalization.
Indeed, the characteristic time scales that rule the dynamics play a central role in determining the asymptotic state of the system. Furthermore, we show that the usual approximations considered in describing the evolution fail to detect the non-Markovianity of the dynamics correctly, and therefore one needs to consider always the full form of the master equation. Finally we show that, in the asymptotic regime, non-Markovianity cannot be detected by the other usual approach based on the non-monotonic behavior of the distance between states evolving in the channel~\cite{PhysRevA.84.052118}. The "state-dependent" characterization of the non-Markovianity of the dynamics must thus be complemented by the "channel-dependent" characterization based on the violation of the divisibility property of the dynamical map.

The paper is organized as follows. In Sec.~\ref{GaussChQBM} we review the basic formalism that describes a Gaussian quantum channel, namely a map that sends Gaussian states into Gaussian states, and the, properly normalized, non-Markovianity measure introduced in~\cite{PhysRevLett.115.070401}. In Sec.~\ref{NMQBMM} we review the Quantum Brownian Motion (QBM) and the pure damping (PD) channels.
In Sec.~\ref{Result} we obtain the explicit expression of the non-Markovianity measure for the QBM channel and show that, due to the structure of the master equation, it is nonvanishing for the entire evolution, regardless of the system-environment interaction, at variance with the case of the PD channel.
Finally, we show that the RWA usually considered in characterizing the system-environment interaction is not capable to capture the non-Markovian property of the dynamics, and therefore is not suitable for a correct description of the dynamics.
Conclusions and outlook are summarized in Sec.~\ref{Conclusion}, together with a comparison between our approach and the one based on the non-monotonicity of the distance between states. In Appendix~\ref{AppQBMRev}, we provide a self-contained review of the QBM channel, together with the asymptotic expressions of the master equation coefficients. Finally, in the technical Appendix~\ref{AppPNMCalc} we provide the detailed calculation of the non-Markovianity measure for the QBM channel.

\section{Non-Markovianity of Gaussian channels: conditions and measures }
\label{GaussChQBM}

In this Section we briefly review the basic mathematical formalism for the description of bosonic Gaussian channels, and the non-Markovianity measure introduced in~\cite{PhysRevLett.115.070401}.

We start by recalling~\cite{RevModPhys.84.621,1751-8121-40-28-S01,Ferraro(2005)} that a state $\rho$ of a CV system with $N$ Bosonic modes admits a representation in terms of the characteristic function:
\begin{equation}\label{Chideffun}
	\chi(\rho)[\Lambda]=\textrm{Tr}[\rho D(\Lambda)] \, ,
\end{equation}
where $D(\Lambda)=\exp[i R^\intercal\Omega\Lambda]$ is the displacement operator, $$\Omega=\bigoplus_{k=1}^N\left(\begin{array}{cc}
0 & 1 \\
-1 & 0
\end{array} \right)$$ is the symplectic matrix, $R=(\hat{x}_1,\hat{p}_1,\ldots,\hat{x}_N,\hat{p}_N)^\intercal$, where $\hat{x}_i$, $\hat{p}_i,i=1,\ldots N$, are the quadrature operators, and $\Lambda=(x_1,p_1,\ldots,x_N,p_N)^\intercal$ is the coordinate vector. A state is Gaussian if, by definition, has a Gaussian characteristic function. As a consequence it can be uniquely characterized by its first order moments
(namely the displacement vector) and its covariance matrix $\sigma$.

An $N$-mode Gaussian quantum channel is a map that preserves the Gaussian form of a Gaussian input state. Its action
can be characterized by the following transformation on the covariance matrix of the input state~\cite{GaussianChannel}:
\begin{equation}\label{sigmaevol}
\sigma(t)=X(t)\sigma(0)X(t)^\intercal+Y(t) \, ,
\end{equation}
namely through the two $2N\times 2N$ real matrices $(X,Y)$.

The non-Markovianity of the dynamics can be assessed through the violation of the divisibility condition
of the intermediate dynamics. This approach has been introduced by Rivas, Huelga, and Plenio for finite dimensional systems in Ref.~\cite{PhysRevLett.105.050403}, and subsequently extended to CV Gaussian channels by Torre, Roga, and Illuminati in Ref.~\cite{PhysRevLett.115.070401}.
Considering the system evolution from time $t_0$ to $t_2$, described by the following family of trace-preserving linear maps $\{\Phi(t_2,t_0),t_2\geq t_0\geq 0\}$, the intermediate dynamics for every time $t_1$, with $t_2\geq t_1\geq t_0$, can be expressed as $\Phi(t_2,t_1)=\Phi(t_2,t_0)\Phi^{-1}(t_1,t_0)$. The evolution is non-Markovian if and only if $\Phi(t_2,t_1)$ fails to be completely positive. For ease of notation, we set $t_0=0$, $t_1=t$ and $t_2=t+\epsilon$, for any instance of $t$ and $\epsilon$.
It has been shown that violation of the divisibility property for Gaussian channels is expressed by the condition~\cite{PhysRevLett.115.070401}:
\begin{equation}\label{ZMat}
Z(t+\epsilon,t) \doteq Y(t +\epsilon, t) - \frac{i}{2} \Omega + \frac{i}{2} X(t +\epsilon, t) \Omega X^\intercal (t +\epsilon, t) < 0,
\end{equation}
i.e. the non positivity of the $Z(t+\epsilon,t)$ matrix,
where $ Y(t +\epsilon, t)$ and $ X(t +\epsilon, t) $
are the matrices that define the intermediate dynamics:
\begin{align}
&X(t+\epsilon,t)\!=\!X(t+\epsilon,0)X^{-1}(t,0) \, , \label{QBMX} \\
& \nonumber \\
&Y(t+\epsilon,t)\!=\!Y(t+\epsilon,0)\!-\!X(t+\epsilon,t)Y(t,0)X^\intercal(t+\epsilon,t) \, . \label{QBMY}
\end{align}
The non-Markovianity of the time evolution can then be quantified by the extent to which the $Z$ matrix in Eq.~(\ref{ZMat}) fails to be positive.
For an $N$-mode Gaussian channel, an immediate choice of the measure is then the punctual non-Markovianity $\mathcal{N}_p$ that quantifies the degree of non-Markovianity at a specific given time $t\in [0,\infty)$~\cite{PhysRevLett.115.070401}:
\begin{equation}\label{Fmeas}
\mathcal{N}_p(t) \doteq \lim_{\epsilon\rightarrow 0^+}\dfrac{\mu[Z(t+\epsilon,t)]}{\nu[Z(t+\epsilon,t)]} \, ,
\end{equation}
where
$\mu[Z(t+\epsilon,t)]$ and $\nu[Z(t+\epsilon,t)]$ are, respectively, the negative part of the spectrum and the sum of the absolute values of the eigenvalues of the $2N$-dimensional matrix $Z(t+\epsilon,t)$:
\begin{align}
&\mu[Z(t+\epsilon,t)] = \dfrac{1}{2}\sum_{i=1}^{2N}\left(\!\vphantom{\dfrac{1}{2}} \vert\lambda_i(t+\epsilon,t)\vert-\lambda_i(t+\epsilon,t)\!\right) \, , \label{mumeas} \\
&\nu[Z(t+\epsilon,t)] = \sum_{i=1}^{2N}\vert\lambda_i(t+\epsilon,t)\vert \, . \label{numeas}
\end{align}
From these definitions it follows that the punctual non-Markovianity Eq.~(\ref{Fmeas}) is positive semi-definite ($\mathcal{N}_p(t)\geq 0$), being zero if and only if the negative part of the spectrum
is zero ($\mu[Z(t+\epsilon,t)]=0$), corresponding to the Markovian case. Furthermore it is normalized in the unit interval ($0\leq \mathcal{N}_p(t)\leq 1$), with maximal non-Markovianity $\mathcal{N}_p(t) = 1$ corresponding to an entirely negative spectrum of the $Z$ matrix.

\section{Quantum Brownian motion and Pure damping channels}
\label{NMQBMM}

In this Section we review the two paradigmatic Gaussian quantum channels, the Quantum Brownian Motion (QBM) and the Pure Damping (PD) channels, whose non-Markovianity will be investigated in the following sections. We will see that the interplay between damping and diffusion effects is of fundamental importance in characterizing the non-Markovianity of the exact dynamics, especially in the asymptotic regime, for different system-environment interaction models.

\subsection{Quantum Brownian Motion}

The QBM Gaussian channel describes the evolution of a quantum harmonic oscillator of frequency $ \omega_0 $ in interaction with $ N $ independent bosonic quantum oscillators that constitute the environment. The associated exact master equation reads~\cite{PhysRevD.45.2843, PhysRevA.70.032113, PhysRevA.67.042108}:

\begin{align}
	\label{master}
	\dot{\rho}(t) &= \! -i[H_0(t), \rho(t)] - \! i\gamma(t)[\hat{x}, \lbrace \hat{p}, \rho(t) \rbrace] + \nonumber \\
 & \nonumber \\
	&- \! \Delta(t)[\hat{x},[\hat{x},\rho(t)]] + \! \Pi(t) [\hat{x},[\hat{p},\rho(t)]] \; .
\end{align}
Here, $ H_0(t) $ is the free Hamiltonian of the system, $ \gamma (t) $ is the damping coefficient, $ \Delta (t) $ and $ \Pi(t) $ are, respectively, the direct and anomalous diffusion coefficients, and $ \hat{x} $ and $ \hat{p} $ are the quadrature operators. Details can be found in Appendix~\ref{AppQBMRev}.

In the rotating wave (RWA) and secular approximations, one neglects the effects due to fast oscillations, i.e. the last term in the r.h.s. of Eq.~(\ref{master}). Resorting to the interaction picture, the master equation in the RWA and secular approximation reads:
\begin{align}\label{masterRWA}
	\dot{\rho} & =\dfrac{\Delta(t)+\gamma(t)}{2}[2 a\rho a^\dagger-a^\dagger a \rho-\rho a^\dagger a]+ \nonumber \\
	& +\dfrac{\Delta(t)-\gamma(t)}{2}[2 a^\dagger\rho a-a a^\dagger \rho-\rho a a^\dagger],
\end{align}
where $a$ and $a^\dagger$ are the Bosonic annihilation and creation operators.

\subsection{Pure Damping channel}

The pure damping (PD) channel is described by the following phenomenological master equation:
\begin{equation}\label{dampingmaster}
\dot{\rho}=\frac{\gamma(t)}{2}[2 a\rho a^\dagger-a^\dagger a \rho-\rho a^\dagger a] \, ,
\end{equation}
which is a particular case of the approximate master equation Eq.~(\ref{masterRWA}) when $\vert\Delta(t)-\gamma(t)\vert\ll\vert\Delta(t)+\gamma(t)\vert$.

\subsection{System-environment interaction}

The explicit solution of Eq.~(\ref{master}), Eq.~(\ref{masterRWA}), and Eq.~(\ref{dampingmaster}), for a fixed initial state, can be obtained once the system-environment coupling, i.e. the spectral density, has been assigned (see Appendix~\ref{AppQBMRev} for details).
In the following we will consider the Ohmic-like class of spectral density distributions:
\begin{equation}
\label{spec}
J_s(\omega)=\left(\frac{\omega}{\omega_c}\right)^s e^{-\frac{\omega}{\omega_{c}}} \; ,
\end{equation}
where $\omega_{c}$ is the cut-off frequency of the bath and $s$ is a parameter characterizing the system-environment interaction. The case $s=1$ corresponds to the Ohmic distribution (i.e. a linear dependence on the frequency for $\omega\ll\omega_c$). For $s<1$ the spectrum is known as sub-Ohmic, for $s>1$ is known as super-Ohmic.

We have investigated the $s=1$ (Ohmic) case and, respectively, the cases $s=1/2$ and $s=3$ for sub-Ohmic and super-Ohmic classes of spectral distributions.
In the following, we will focus on the high-temperature regime, namely the case in which the classical thermal energy is much larger than the typical quantum exchange energy units: $k_B T\gg\hbar \omega_c,\hbar \omega_0$. In this regime, non-Gaussian corrections to the system-bath dynamics can be neglected, and the average number of excitations is essentially linear in the temperature (see Appendix A), so that the dynamics is exactly solvable. Since in the high-temperature regime the three spectral settings yield qualitatively equivalent results, in the following we will report exclusively on the Ohmic case.

\section{Non-Markovianity of Gaussian channels: finite-time and asymptotic regimes}\label{Result}

In this Section we study the non-Markovianity of the channels reviewed in Section~\ref{NMQBMM}, both for the exact and the approximate dynamics. We will show that the standard approximations lead to an incorrect assessment of the non-Markovianity of Gaussian channels and fail to detect it in the asymptotic regime.

\subsection{Non-Markovianity of the QBM channel}

The $ 2N \times 2N $ matrices $ (X,Y) $ that characterize the exact evolution Eq.~(\ref{master}) are (see Appendix~\ref{AppQBMRev}):
\begin{align}
\label{XY}
&X(t) = e^{-\frac{\Gamma(t)}{2}} R(t) \, , \nonumber \\
& \nonumber \\
&Y(t) = 2 \bar{W}(t) \, ,
\end{align}
where $\Gamma(t)$ is defined in terms of the damping coefficient as $\Gamma(t)=2\int_0^t\gamma(s)ds$, $R(t)$ is the rotation matrix Eq.~(\ref{WBarMat}), and $\bar{W}(t)$ is given in Eq.~(\ref{WBAR}).

The eigenvalues $\lambda_{\pm}$ of the $Z(t+\epsilon,t)$ matrix Eq.~(\ref{ZMat}) are obtained from Eqs.~(\ref{XY}) through the matrices Eqs.~(\ref{QBMX}) and~(\ref{QBMY}). The details are given in Appendix~\ref{AppPNMCalc}. One has:
\begin{equation}\label{eigenvalues}
\lambda_{\pm}(t) = \Delta(t) \pm \sqrt{\Delta(t)^2 + \gamma(t)^2 + \Pi(t)^2} \, .
\end{equation}

Provided that the direct diffusion coefficient $\Delta(t)$ is non negative at all times, the eigenvalue $\lambda_+(t)$ is always positive. On the other hand, the eigenvalue $\lambda_-(t)$ is certainly negative, and thus the time evolution is certainly non-Markovian, provided that either the damping coefficient $\gamma(t)$ or the anomalous diffusion coefficient $\Pi(t)$, or both, are nonvanishing. In fact, by evaluating the explicit expressions of the master equation coefficients in the exact master equation for the QBM, one finds that $\lambda_-(t)$ is negative at all times~\cite{PhysRevA.80.062324}.

\begin{figure}[t]
	\centering
	\includegraphics[width=8.5cm]{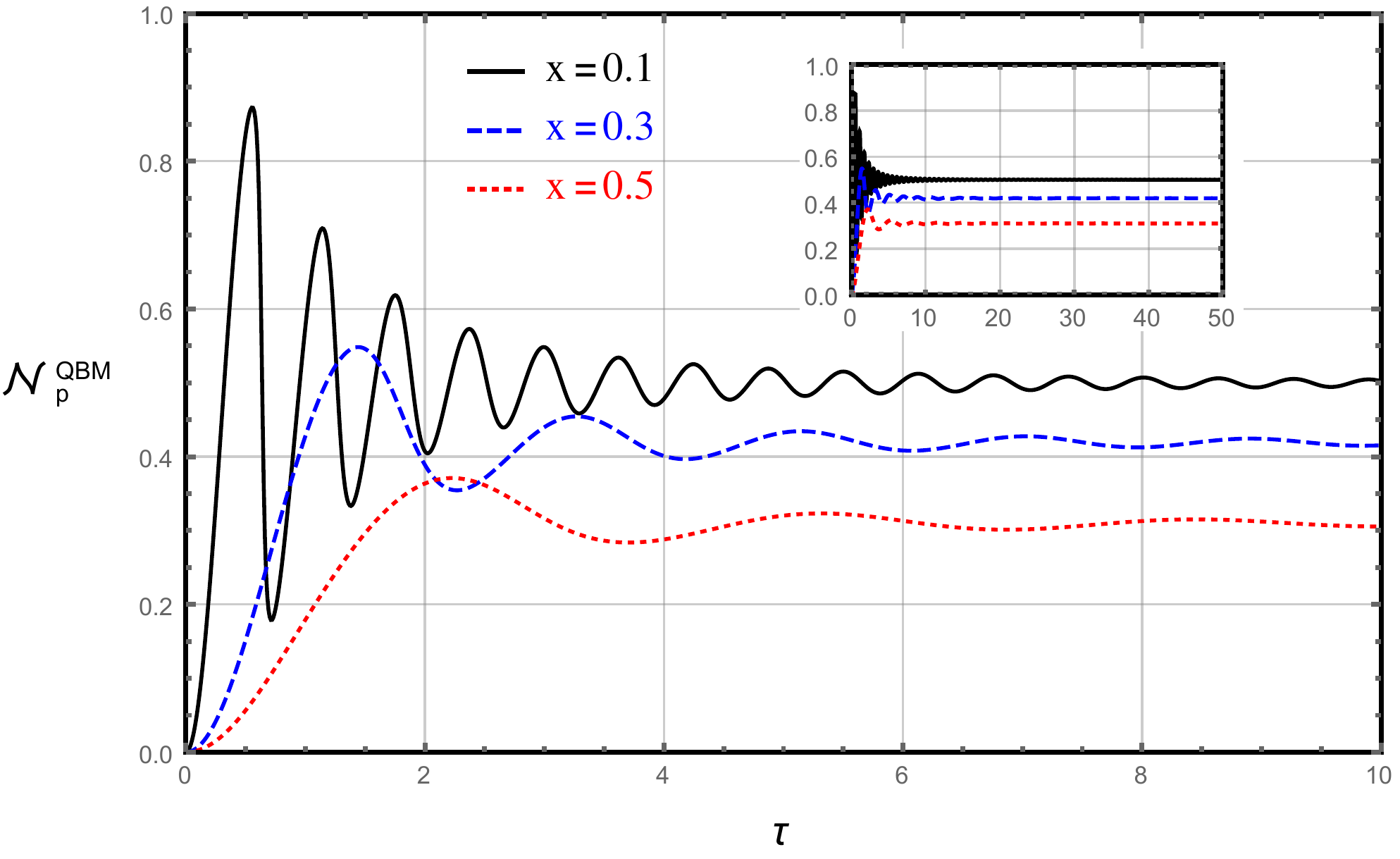}
	\caption{(color online) Punctual non-Markovianity Eq.~(\ref{PNMARKEXP}) for the QBM channel under the exact dynamics Eq.~(\ref{master}), for a non-Markovianity parameter $x = 0.1$ (black full line), $x = 0.3$ (blue dashed line), and $x = 0.5$ (red dotted line), as a function of the dimensionless time $\tau=\omega_c t$, in the high-temperature regime $\frac{k_B T}{\hbar \omega_c}=100$, so that $x\gtrsim 0.1$ (see Appendix~\ref{AppQBMRev}). The behaviour of $\mathcal{N}_p^{QBM}$ in the asymptotic regime is reported in the inset. Asymptotically, the punctual non-Markovianity tends to a constant whose numerical value is determined by the strength of the $x$ parameter. All plotted quantities are dimensionless.}
	\label{eigen}
\end{figure}

It is useful to express all quantities in terms of the dimensionless time $\tau=\omega_c t$ and the non-Markovianity parameter, namely the ratio $ x = \tau_R / \tau_E = \omega_c / \omega_0 $ between the correlation time scale $ \tau_E $ of the environment and the relaxation time scale $ \tau_R $, that corresponds to the rate of the system state change due to the system-environment interaction~\cite{PhysRevA.70.032113,PhysRevA.80.062324}.
Through Eqs.~(\ref{mumeas}) and~(\ref{numeas}), the punctual non-Markovianity Eq.~(\ref{Fmeas}) for the exact dynamics of the QBM channel, Eq.~(\ref{master}), reads:
\begin{equation}\label{PNMARKEXP}
\mathcal{N}_p^{QBM}(\tau, x)\!=\!\dfrac{1}{2}\left[1-\dfrac{\Delta(\tau, x)}{\sqrt{\Delta(\tau, x)^2+\gamma(\tau, x)^2+\Pi(\tau, x)^2}}\right]\!\!.
\end{equation}

The behaviour of the punctual non-Markovianity, Eq.~(\ref{PNMARKEXP}), as a function of $\tau$ for the Ohmic case and for different values of the $x$ parameter is reported in Fig.~\ref{eigen}. Its behaviour in the asymptotic time regime is reported in the inset.
The Markovian regime is recovered in the limit $x\rightarrow \infty$: in this limit the diffusion coefficient $\Delta(\tau, x)$ diverges at all times~\cite{PhysRevA.70.032113} and $\mathcal{N}_p^{QBM}(\tau) \rightarrow 0$ at all times.

In order to analyze the behavior of Eq.~(\ref{PNMARKEXP})
in the long-time regime $\tau\gg 1$, we consider the explicit asymptotic limit $\tau\rightarrow \infty$ of Eq.~(\ref{PNMARKEXP}) for which the expressions of the master-equation coefficients take a simpler form (see Appendix~\ref{AppQBMRev} for details).
Through Eq.s~(\ref{gammainf}),~(\ref{deltainf}), and~(\ref{piinf}) we have:
\begin{align}\label{AsNMx}
&\mathcal{N}_{p,asymp}^{QBM}(x) \doteq \lim_{\tau\rightarrow\infty}\mathcal{N}_p^{QBM}(\tau, x)=\frac{1}{2} + \nonumber \\
& \nonumber \\
&-\frac{k_B T\pi x}{\hbar \omega_c \sqrt{\dfrac{4 k_B^2 T^2 x^2}{\hbar^2 \omega_c^2 } \left[\left(\text{Ei}\left(\frac{1}{x}\right)\!-\!e^{2/x} \text{Ei}\left(-\frac{1}{x}\right)\right)^2+\pi ^2\right]+\pi ^2}} \, .
\end{align}
Here $\text{Ei}\left(x\right)$ is the exponential integral function~\cite{Luke1969}.
From Eq.~(\ref{AsNMx}) it follows that
$\mathcal{N}_{p,asymp}^{QBM}(x)>0$ for every finite value of the $x$ parameter. The behaviour of the punctual non-Markovianity in the long-time regime is reported in the inset of Fig.~\ref{eigen}.

  \begin{figure}[t]
  	\centering
  	\includegraphics[width=8.5cm]{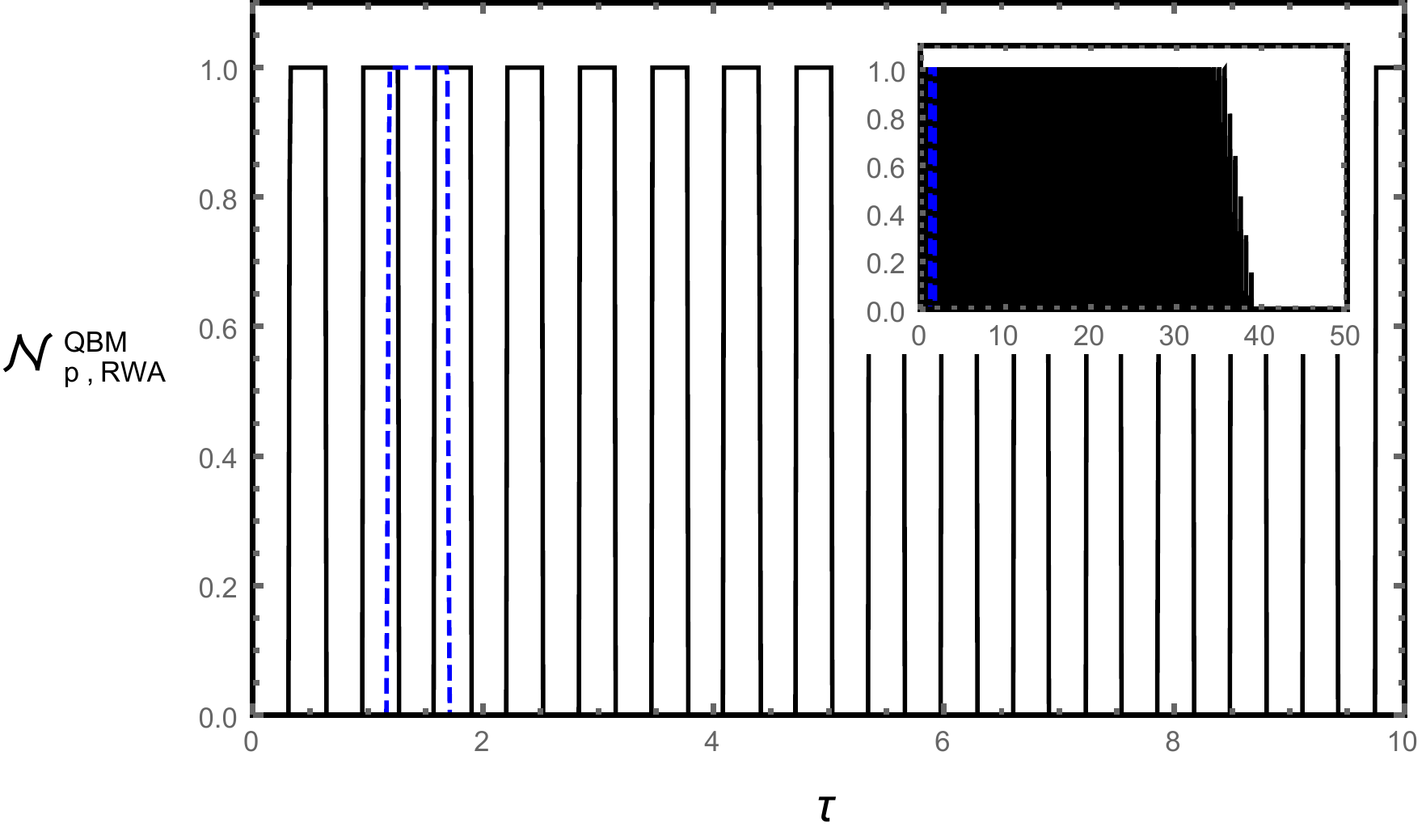}
  	\caption{(color online) Punctual non-Markovianity Eq.~(\ref{NMarkMeasRWA}) for the QBM channel within the RWA approximation to the dynamics, Eq.~(\ref{masterRWA}), for a non-Markovianity parameter $x = 0.1$ (full black line) and  $x = 0.3$ (blue dashed line), as a function of the dimensionless time $\tau=\omega_c t$, in the high-temperature regime $\frac{k_B T}{\hbar \omega_c}=100$, so that $x\gtrsim 0.1$ (see Appendix~\ref{AppQBMRev}). At variance with the exact dynamics, in the RWA approximation there are finitely many finite time intervals in which the channel is Markovian at any finite $x$. Furthermore, within the RWA, the QBM channel is always Markovian in the asymptotic regime. As shown in the inset, $\mathcal{N}_{p,\textrm{RWA}}^{QBM}$ vanishes identically at sufficiently large times. All plotted quantities are dimensionless.}
  	\label{RWAPIC}
  \end{figure}

We now consider a different measure, the integrated one, that quantifies the total amount of non-Markovianity along the entire dynamical evolution. For any time interval $I\subseteq [0,\infty)$ it is defined as:
\begin{equation}\label{NMarkMeasInt}
\mathcal{N}_{\textrm{int}}(x) \doteq \dfrac{\int_I \mathcal{N}_p(\tau, x) d\tau}{\int_If[\mathcal{N}_p(\tau, x)]d\tau} \; ,
\end{equation}
where the function $f(y)$ is defined as:
\begin{equation}
f(y) \doteq \left\lbrace
\begin{array}{l}
0 \textrm{   if   }y=0 \, ,\\
1 \textrm{   otherwise \, . }
\end{array}
\right.
\end{equation}
The denominator of Eq.~(\ref{NMarkMeasInt}) represents the sum of the lengths of the intervals in which the evolution is punctually non-Markovian.

It is important to observe that, for the QBM channel, the punctual non-Markovianity coincides, in the asymptotic limit, with the integrated one.
Indeed, for the entire dynamics, we have:
\begin{align}
\mathcal{N}&_{\textrm{int}}^{QBM}(x) \doteq \dfrac{\int_0^\infty \mathcal{N}_p^{QBM}(\tau, x) d\tau}{\int_0^\infty f[\mathcal{N}_p^{QBM}(\tau, x)]d\tau}= \nonumber \\
& \nonumber \\
&\lim_{a\rightarrow\infty}\dfrac{\int_0^a \mathcal{N}_p^{QBM}(\tau, x) d\tau}{\int_0^a f[\mathcal{N}_p^{QBM}(\tau, x)]d\tau}
= \nonumber \\
& \nonumber \\
& \dfrac{\mathcal{N}_{p,asymp}^{QBM}(x)}{f[\mathcal{N}_{p,asymp}^{QBM}(x)]} = \mathcal{N}_{p,asymp}^{QBM}(x) \, ,
\end{align}
where we have applied the De L'H\^{o}pital rule and the fact that $f[\mathcal{N}_{p,as}^{QBM}(x)] = 1$. As a consequence, the measure of asymptotic non-Markovianity is uniquely defined as follows:
\begin{equation}
\mathcal{N}_{p,asymp}^{QBM}(x) = \mathcal{N}_{\textrm{int}}^{QBM}(x) \doteq \mathcal{N}_{asymp}^{QBM}(x) \, .
\end{equation}

\begin{table*}[!t]
	\caption{\label{SummTable} Comparison between different non-Markovianity quantifiers for the QBM channel in the finite and asymptotic time regimes.}
	\begin{ruledtabular}
		\begin{tabular}{ccc}
			Non-Markovianity quantifier & Finite-time regime & Asymptotic regime \\
			\hline \\
			\phantom{a} & There may exist intervals of time & \phantom{a} \\
			State-based witness & for which the dynamics & There is no residual non-Markovianity \\
			\phantom{a}	 &  is Markovian & \phantom{a}  \\ \\
			\phantom{a} & The exact dynamics (no RWA)  & The asymptotic non-Markovianity  \\
			Channel-based measure & is non-Markovian at any finite time & is non vanishing for any \\
			\phantom{a} & for any finite $x$ & finite $x$ \\\\
		\end{tabular}
	\end{ruledtabular}
\end{table*}

\subsection{Non-Markovianity of the QBM channel within the RWA approximation}

Finally, we want to asses whether and how the use of the RWA approximation affects the correct evaluation of the non-Markovianity. Starting from the approximated form of the $(X,Y)$ matrices Eqs.~(\ref{XY}) that describe the evolution~\cite{PhysRevA.84.052118,PhysRevA.80.062324}, it is straightforward to obtain the following expression for the punctual non-Markovianity in the RWA approximation:
\begin{equation}
\mathcal{N}_{p,\textrm{RWA}}^{QBM}(\tau, x) \!=\!\dfrac{1}{2}\!\left[\!1\!-\!\dfrac{2\Delta(\!\tau\!, \!x\!)}{\vert\Delta(\!\tau\!, \!x\!)\!+\!\gamma(\!\tau\!, \!x\!)\vert+\vert\Delta(\!\tau\!, \!x\!)\!-\!\gamma(\!\tau\!, \!x\!)\vert}\right], \label{NMarkMeasRWA} \\ 	
\end{equation}
In the high-temperature regime, the condition $\Delta(\tau, x)\gg\gamma(\tau, x)$ is certainly satisfied. Consequently, the value of the punctual non-Markovianity, Eq.~(\ref{NMarkMeasRWA}), depends essentially on the sign of the diffusion coefficient $\Delta(\tau, x)$. This is reflected in the binary structure of the punctual non-Markovianity as a function of time, as reported in Fig.~\ref{RWAPIC}. Indeed, from the explicit expression, Eq.~(\ref{NMarkMeasRWA}), and from the spectral densities considered, one verifies directly that there exist finitely many finite intervals of time for which $\Delta(\tau, x)>0$. In these intervals $\mathcal{N}_{p,\textrm{RWA}}^{QBM}(\tau, x) = 0$, and the channel is Markovian. Furthermore, since $\lim_{\tau\rightarrow\infty}\Delta(\tau, x)$ is a positive constant, $\mathcal{N}_{p,\textrm{RWA}}^{QBM}$ vanishes identically in the asymptotic regime, as illustrated in the inset of Fig.~\ref{RWAPIC}.

\subsection{Non-Markovianity of the PD channel}

Turning to the PD channel, Eq.~(\ref{dampingmaster}), the non-Markovianity features a behaviour qualitatively similar to that of the QBM channel in the RWA. Indeed, starting from the corresponding approximated form of the $(X,Y)$ matrices Eqs.~(\ref{XY})~\cite{PhysRevA.84.052118,PhysRevA.80.062324}, the punctual non-Markovianity of the PD channel takes the very simple form:
\begin{equation}
	\mathcal{N}_p^{PD}(\tau, x)\!=\!\dfrac{1}{2}\!\left[1-\dfrac{\gamma(\tau, x)}{\vert\gamma(\tau, x)\vert}\right] \label{NMarkMeasDP} \, .
\end{equation}
From Eq.~(\ref{NMarkMeasDP}) we see that the channel dynamics is non-Markovian if and only if the damping coefficient $\gamma(\tau, x)$ is negative. Indeed, for the classes of spectral densities considered, there always exist intervals of time in which the damping coefficient is positive and the dynamics is thus Markovian. Moreover, in the asymptotic regime $\gamma(\tau, x)$ tends to a constant positive value, and the non-Markovianity vanishes identically.

\section{Conclusions and outlook}\label{Conclusion}

In the present work, by exploiting a necessary and sufficient criterion of non-Markovianity based on the violation of the divisibility of the dynamical map, we have shown that the exact dynamics of Gaussian quantum channels can be non-Markovian at all times $t\in (0,\infty)$, as illustrated in the paradigmatic case of the QBM channel. Moreover, we have verified that the approximations usually considered in describing the system dynamics, such as the RWA, fail in general to  preserve and assess correctly the non-Markovian character of the time evolution both in the finite-time and asymptotic regimes. A nonvanishing non-Markovianity in the asymptotic regime leads to some profound consequences: although at very large times the state of the system does not depend anymore on the initial input state, the asymptotic output still depends on the bath configurations via the time scales that rule the open system dynamics being considered.

In assessing the non-Markovianity, it is interesting to compare the approach followed in the present work with the approach based on the
distinguishability between states evolving in the channel~\cite{PhysRevLett.103.210401,RevModPhys.88.021002}.
Indeed, since in a Markovian channel the destructive effect of the system-environment interaction makes two different input states less distinguishable, the memory effects of the dynamics can then be checked through an increasing of their distinguishability (information backflow from the environment to the system). This approach was introduced in Ref.~\cite{PhysRevLett.103.210401} for the finite-dimensional case, and subsequently extended to the CV setting in Ref.~\cite{PhysRevA.84.052118}. If we resort to a distance measure between quantum states that has the property to be contractive under trace-preserving and completely positive maps, one can introduce in analogy with Eq.~(\ref{Fmeas}) the following distance-based quantifier of punctual non-Markovianity:

\begin{equation}\label{DistMeas}
	\mathcal{N}^D_p(t)\equiv\max\{0,\min_{\rho_1,\rho_2}\dfrac{d}{dt}D[\rho_1(t),\rho_2(t)]\} \, ,
\end{equation}
where the minimization must be taken over the entire set of all possible input states.
When the channel is Markovian, the time-derivative is negative (the states become less distinguishable) and the non-Markovianity
vanishes identically. It is known that this approach leads only to a sufficient condition for a quantum channel to be non-Markovian, and the quantifier
Eq.~({\ref{DistMeas}}) is thus, strictly speaking, not a measure of non-Markovianity, but rather a non-Markovianity witness~\cite{0034-4885-77-9-094001}. In passing, we remark that, at variance with the geometric, distance-based approach, the approach based on the violation of the divisibility does not require a complex maximization procedure, and is thus computationally efficient. 
	
In the following we discuss how, in the asymptotic regime, the quantifier Eq.~({\ref{DistMeas}}) fails to detect the asymptotic non-Markovianity of the QBM dynamics. Resorting to the characteristic function representation of the input states Eq.~(\ref{Chideffun}), its evolution in the QBM channel is given by Eq.~(\ref{chiEvol}). In the asymptotic limit, the damping coefficient Eq.~(\ref{gammainf}) assumes a constant positive value and therefore
$ \lim_{t \rightarrow \infty} \Gamma(t) =2 \lim_{t \rightarrow \infty}\int_0^t\gamma(s)ds= \infty $. Moreover, since
$ \chi_0 (\mathbf{0}) = 1 $, where $ \mathbf{0} = (0,0)^\intercal $ is the null vector~\cite{Barnett&Radmore}, it is straightforward to show from Eq.~(\ref{chiEvol}) that, for every initial input state:
\begin{equation}
\label{chiasymp}
\chi_{asymp} (\Lambda) \equiv \lim_{t \rightarrow \infty} \chi_t (\Lambda) = e^{-\Lambda^\intercal \bar{W}_{asymp}\Lambda} \; ,
\end{equation}
where $ \bar{W}_{asymp}$ is the asymptotic value of the $ \bar{W}(t) $ matrix, Eq.~(\ref{WBAR}).
Hence, the asymptotic state is always the same state, irrespective of the choice of the initial input state. As a consequence, in the asymptotic regime, the time-derivative in Eq.~(\ref{DistMeas}) is negative and $\mathcal{N}_{p,asymp}^D=0$.

In conclusion, the non-Markovianity properties of a Gaussian channel in the asymptotic regime can be assessed correctly only resorting to measures directly relying on the channel structure, as in the case of the non-Markovianity measure based on the violation of the divisibility of the dynamical map.

In the distance-based approach to the quantification of non-Markovianity, the intuitive interpretation in terms of information flux from the environment back to the system is highlighted by the "re-coherence" effect, represented by an increased distinguishability between states. The authors in Ref.~\cite{PhysRevA.93.042119} show, for a particular channel, that this information flux can be connected to an increase in the quantum correlations, and that the violation of the divisibility property of the dynamical map does not correspond to this increase. It therefore represents only a necessary condition to obtain a backflow of information.

On the other hand, that the approach based on the violation of the divisibility property of the dynamical map in fact
highlights different aspects of the non-Markovianity appears clearly from the work reported in Ref.~\cite{Groblacher(2015)}. In that paper, it is shown that the memory of the evolution based on the violation of the divisibility of the dynamical map allows to reconstruct the spectral density of the bath, and that this information is not recoverable resorting to the distance-based approach, as highlighted by Eq.~(\ref{chiasymp}). The comparison between the two approaches is summarized in Tab.~\ref{SummTable}. The two approaches are clearly complementary. State-based quantifiers are only witnesses and require a complex optimization on the class of input states. Channel-based quantifiers are proper measures and are computationally efficient. On the other hand, given that it is in general very hard to solve exactly the complete dynamics of an open quantum system, state-based witnesses can still be useful and more relevant any time one needs to resort to approximate forms of the dynamical evolution. In such instances one would need to use and compare carefully both approaches in order to extract useful information.

More generically, characterizing and quantifying the non-Markovianity of quantum channels via the violation of the divisibility property of the dynamical map relates this dynamical feature to an intrinsic property of the evolution that is not directly related to the dynamics of the system-environment quantum correlations of the states evolving in the channel. 

Such an intrinsic characterization of non-Markovianity might thus lead to the identification of resources for quantum technologies that are "channel-based" rather than "state-based". It would thus be interesting to assess the relevance of the non-Markovianity quantification provided by Eq.~(\ref{Fmeas}) in the context of quantum information processing.

In particular, it would be worth investigating whether and how the non-Markovianity of Gaussian channels might be exploited to improve the performance of CV quantum information protocols. Indeed, work is in progress along these directions in order to verify the possibility of improved, non-Markovianity assisted CV quantum teleportation protocols in realistic conditions, exploiting the exact results obtained in the present paper on the dynamics of the QBM channel~\cite{Torre2018}.

\appendix

\section{Quantum Brownian Motion channel}\label{AppQBMRev}

In this Section we give a brief review of the exact master equation for the QBM channel, Eq.~(\ref{master}), and of its solutions. The full details are reported in Refs.~\cite{PhysRevD.45.2843, PhysRevA.67.042108, PhysRevA.70.032113}.

Quantum Brownian motion describes the evolution of a quantum mechanical oscillator, characterized by a frequency $\omega_0$, in contact with a bath of harmonic oscillators via a position-position coupling. We focus on the particular case of factorized initial conditions.
We recall that the system evolution is described by the master equation Eq.~(\ref{master}):
\begin{align*}
\frac{d\rho}{dt} = & -\frac{i}{\hbar} [H_0,\rho(t)] - \Delta(t)[\hat{x},[\hat{x},\rho(t)]] + \\
& + \Pi(t)[\hat{x},[\hat{p},\rho(t)]] -i \gamma(t)[\hat{x},\{\hat{p},\rho(t)\}] \; ,
\end{align*}
where $H_0$ is the free Hamiltonian of the system, $\Delta(t)$ and $\Pi(t)$ are, respectively, the normal and anomalous diffusion coefficients, $\gamma(t)$ is the damping coefficient, and $\hat{x}$ and $\hat{p}$ are the quadrature operators.

The coefficients read~\cite{PhysRevA.67.042108}:
\begin{equation}
	\gamma(t)\!=\!\alpha^2\!\int_0^t  \, ds\int_0^{+\infty} \, d\omega J (\omega)  \sin (\text{$\omega $s}) \sin (\text{$\omega_{0} $s}),\label{gamma}
\end{equation}
\begin{equation}
\Delta(t)\!=\!\alpha^2\!\int_0^t  \,\!\! ds\int_0^{+\infty} \, \!\!\!\!\!\!\!\! d\omega J (\omega) (2N(\omega)+1) \cos (\text{$\omega $s}) \cos (\text{$\omega_{0} $s}),\label{delta}
\end{equation}
\begin{equation}
\Pi(t)\!=\!\alpha^2\!\int_0^t  \,\!\! ds\int_0^{+\infty} \, \!\!\!\!\!\!\!\! d\omega J (\omega) (2N(\omega)+1) \cos (\text{$\omega $s}) \sin (\text{$\omega_{0} $s}),\label{pi}
\end{equation}
where $\alpha$ is the oscillator-bath coupling constant, $N(\omega)=[\exp(\hbar \omega/k_B T)-1]^{-1}$ is the mean number of photons, $J(\omega)$ is the spectral density, that models the system-environment interaction,
and $\omega_c$ is the cut-off frequency of the environment.

In the hight-temperature regime, the classical thermal energy is much larger than the typical ones that characterize the system evolution ($k_B T \gg \hbar \omega_c, \hbar \omega_0$). Under this condition we can set $2N(\omega)+1\approx \frac{2k_B T}{\hbar \omega}$ in Eqs.~(\ref{gamma})--(\ref{pi}); consequently it is possible to obtain an explicit expression of the master equation coefficients~\cite{PhysRevA.80.062324}. We note that this condition imposes the constraint $x\gtrsim 0.1$ on the non-Markovianity parameter $x=\frac{\omega_c}{\omega_0}$~\cite{PhysRevA.80.062324}.

In the following, we resort to the phase space formulation of quantum mechanics.
In the characteristic function description, the solution of the master equation Eq.~(\ref{master}) is~\cite{PhysRevA.67.042108, PhysRevA.70.032113}:
\begin{equation}\label{chiEvol}
\chi(\Lambda, t)=\chi(e^{-\frac{\Gamma(t)}{2}}R^{-1}(t)\Lambda, 0)e^{-\Lambda^\intercal \bar{W}(t)\Lambda},
\end{equation}
where:
\begin{align}
&\bar{W}(t)\!=\![R^{-1}(t)]^\intercal\!\!\left[e^{-\Gamma(t)}\!\!\!\int_0^t \!\!ds e^{\Gamma(s)}R^\intercal(s) M(s) R(s)\right]\!\!R^{-1}(t), \label{WBAR}  \\
& M(s)\!=\!\left(\begin{array}{cc}
\Delta(s)& -\Pi(s)/2 \\
-\Pi(s)/2 & 0
\end{array} \right) \; , \\
& \nonumber \\
& R(t)=\left(\begin{array}{cc}
\cos(\omega_0 t)& \sin(\omega_0 t) \\
-\sin(\omega_0 t)& \cos(\omega_0 t)
\end{array} \right) \; ,\label{WBarMat}
\end{align}
with:
\begin{equation}
\Gamma(t)=2\int_0^t \gamma(s)ds \; .
\label{Gammona}
\end{equation}
The Gaussian nature of the evolution is manifestly evident from Eq.~(\ref{chiEvol}): a Gaussian initial characteristic function maintains its Gaussian character during the entire dynamics.

We can now derive the $2\times 2$ $(X,Y)$ matrices that characterize the QBM channel. Considering the case of a generic Gaussian input state, by Eq.~(\ref{chiEvol}) we find the corresponding transformation on its covariance matrix:
\begin{equation}\label{SigmaEvolQBM}
	\sigma(t)=\left[e^{-\frac{\Gamma(t)}{2}}R^{-1}(t)\right]^\intercal\sigma(0)\left[e^{-\frac{\Gamma(t)}{2}}R^{-1}(t)\right]+2\bar{W}(t),
\end{equation}
where $R(t)$ is the rotation matrix Eq.~(\ref{WBarMat}) and $\bar{W}(t)$ is defined in Eq.~(\ref{WBAR}), with $\Gamma(t)$ expressed by Eq.~(\ref{Gammona}).
From Eq.~(\ref{SigmaEvolQBM}) and Eq.~(\ref{sigmaevol}) it is straightforward to obtain Eqs.~(\ref{XY}).

Finally, we report the expression of the master equations coefficients Eq.s~(\ref{gamma}) to~(\ref{pi}) in the asymptotic regime $t \rightarrow \infty$.

For an Ohmic spectral density ($s=1$ in Eq.~(\ref{spec})) and in the high-temperature regime, they read:
\begin{align}
\gamma_{as}(T)&=\frac{1}{2} \pi  \alpha ^2 \omega_0 e^{-\frac{1}{x}}, \label{gammainf} \\
& \nonumber \\
\Delta_{as}(T)&= \frac{\pi  \alpha ^2  k_B T}{\hbar } e^{-\frac{1}{x}}, \label{deltainf} \\
& \nonumber \\
\Pi_{as}(T)&= \frac{2\alpha ^2 k_B T}{\hbar }  \left[ \text{Shi}\left(\frac{1}{x}\right) \cosh \left(\frac{1}{x}\right) + \right. \nonumber \\
&\left.-\text{Chi}\left(\frac{1}{x}\right) \sinh \left(\frac{1}{x}\right) \right],\label{piinf}
\end{align}
where $ \text{Shi}(x)$ and $\text{Chi}(x)$ are respectively the hyperbolic sine integral and hyperbolic cosine integral functions~\cite{Luke1969}, and $k_B$ is the Boltzmann constant.

\section{Calculation of the eigenvalues of the $Z(t+\epsilon,t)$ matrix Eq.~(\ref{ZMat}) for the QBM channel}\label{AppPNMCalc}

In this Section we present the details of the calculation of the eigenvalues, Eq.s~(\ref{eigenvalues}), of the $Z(t+\epsilon,t)$ matrix, Eq.~(\ref{ZMat}), for the QBM channel.

We start from the explicit expression of the $(X,Y)$ matrices. From Eqs.~(\ref{QBMX}),~(\ref{QBMY}), and~(\ref{XY}) we have:
\begin{align}
	X(t+\epsilon,t)&=e^{-\frac{\Gamma(t+\epsilon,t)}{2}}R(\epsilon) \label{XINTMAT}\\
& \nonumber \\
	Y(t+\epsilon,t)&=2\bar{W}(t+\epsilon)-e^{-\frac{\Gamma(t+\epsilon,t)}{2}}R(\epsilon)\bar{W}(t)R^\intercal(\epsilon)\label{YINTMAT},
\end{align}
where $R$ and $\bar{W}$ are respectively the matrices Eq.~(\ref{WBarMat}) and Eq.~(\ref{WBAR}), and where we have defined $\Gamma(t+\epsilon,t)=2\int_t^{t+\epsilon}\gamma(s)ds$.

We can now obtain the expression of the $Z$ matrix, Eq.~(\ref{ZMat}), whose eigenvalues characterize the non-Markovian property of the dynamics.

From Eq.s~(\ref{ZMat}),~(\ref{XINTMAT}), and~(\ref{YINTMAT}), we obtain:
\begin{widetext}
	\begin{equation}\label{ZMATQBM}
	Z(t+\epsilon,t)=2\bar{W}(t+\epsilon)-2e^{-\Gamma(t+\epsilon,t)}R(\epsilon)\bar{W}(t)R^\intercal(\epsilon)-\dfrac{i}{2}\Omega\left[1-e^{-\Gamma(t+\epsilon,t)}\right].
	\end{equation}
\end{widetext}
Due to the condition $\epsilon\ll 1$ we can expand in series Eq.~(\ref{ZMATQBM}) up to the first order in $\epsilon$. We note that:
\begin{widetext}
	\begin{align}
	&2\bar{W}(t+\epsilon)\approx 2\bar{W}(t)+\left[2\Delta(t)J^{00}-\Pi(t)J_2\right]\epsilon
-4\gamma(t)\bar{W}(t)\epsilon+2\omega_0\left[\Omega\bar{W}(t)-\bar{W}(t)\Omega\right]\epsilon \label{P1} \\
	& \nonumber \\ &-2e^{-\Gamma(t+\epsilon,t)}R(\epsilon)\bar{W}(r)R^\intercal(\epsilon)\approx-2\bar{W}(t)-2\omega_0\left[\Omega\bar{W}(t)-\bar{W}(t)\Omega\right]\epsilon+4\gamma(t)\bar{W}(t)\epsilon \label{P2} \\
& \nonumber \\
	&-\dfrac{i}{2}\Omega\left[1-e^{-\Gamma(t+\epsilon,t)}\right]\approx-i\Omega\gamma(t)\epsilon\label{P3},
	\end{align}	
\end{widetext}
where $J^{00}$ and $J_2$ are, respectively, the single-entry matrix and the exchange matrix:
\begin{equation}
		J^{00}=\left(\begin{array}{cc}
		1	& 0 \\
		0	& 0
		\end{array} \right),
\end{equation}
\begin{equation}
			J_2=\left(\begin{array}{cc}
			0	& 1 \\
			1	& 0
			\end{array} \right).
\end{equation}
From Eq.s~(\ref{P1}),~(\ref{P2}), and~(\ref{P3}), Eq.~(\ref{ZMATQBM}) reduces to:

\begin{equation}\label{ZMATFIN}
	Z(t+\epsilon,t)\approx\left[2\Delta(t)J^{00}-\Pi(t)J_2-i\Omega\gamma(t)\right]\epsilon.
\end{equation}

The matrix Eq.~(\ref{ZMATFIN}) represents the intermediate evolution of the system. Its eigenvalues are given by Eq.~(\ref{eigenvalues}).

\bibliography{bibliography}
\bibliographystyle{apsrev4-1}

\end{document}